\documentclass[aps,prb,reprint,superscriptaddress,showpacs,floatfix]{revtex4-1}

\usepackage[dvipdfmx]{graphicx}
\usepackage{mathrsfs}
\usepackage[mathscr]{euscript}
\usepackage{amsmath,amssymb}
\usepackage{bm}
\usepackage{color}
\usepackage[vcentermath]{youngtab}
\usepackage{ulem}

\usepackage[dvipdfmx,
colorlinks=true,
urlcolor=blue,
citecolor=blue,
linkcolor=blue,
hyperfootnotes=false]{hyperref}

\def\Tr{{\rm Tr}}
\def\ket#1{ | #1 \rangle }
\def\bra#1{ \langle #1 | }

\newcommand{\1}{\mbox{1}\hspace{-0.25em}\mbox{l}} %

\begin{document}


\title{Symmetry protected topological phases in two-orbital SU(4) fermionic atoms}

\author{Hiroshi Ueda}
\affiliation{Computational Materials Science Research Team, RIKEN Center for Computational Science (R-CCS), Kobe, Hyogo 650-0047, Japan}
\author{Takahiro Morimoto}
\affiliation{Department of Physics, University of California, Berkeley, California 94720, USA}
\author{Tsutomu Momoi}
\affiliation{Condensed Matter Theory Laboratory, RIKEN Cluster for Pioneering Research (CPR), Wako, Saitama, 351-0198, Japan}
\affiliation{Quantum Matter Theory Research Team, RIKEN Center for Emergent Matter Science (CEMS), Wako, Saitama, 351-0198, Japan}

\date{\today}

\begin{abstract}
We study one-dimensional systems of two-orbital SU(4) fermionic cold atoms. In particular, we focus on an SU(4) spin model [named SU(4) $e$-$g$ spin model] that is realized in
a low-energy state in the Mott insulator phase at the filling $n_g=3, n_e=1$ ($n_g, n_e$: numbers of atoms in ground and excited states, respectively).
Our numerical study with the infinite-size density matrix renormalization group shows that the ground state of SU(4) $e$-$g$ spin model
is  a nontrivial symmetry protected topological (SPT) phase protected by $Z_4 \times Z_4$ symmetry.
Specifically, we find that the ground state belongs to an SPT phase with the topological index $2\in\mathbb{Z}_4$ and show sixfold degenerate edge states.
This is topologically distinct from SPT phases with the index $1\in\mathbb{Z}_4$ that are realized in the SU(4) bilinear model
and  the SU(4) Affleck-Kennedy-Lieb-Tasaki (AKLT) model.
We explore the phase diagram of SU(4) spin models including $e$-$g$ spin model, bilinear-biquadratic model, and AKLT model, and identify that antisymmetrization effect in neighboring spins (that we quantify with Casimir operators) is the driving force of the phase transitions.
Furthermore, we demonstrate by using the matrix product state how the $\mathbb{Z}_4$ SPT state with six edge states appears in the SU(4) $e$-$g$ spin model.
\end{abstract}

\maketitle
\section{Introduction}

Topological phases are phases of matter that are characterized by
topological natures of wave functions instead of symmetry breaking, and they are intensively studied in recent condensed matter research.\cite{Hasan_RMP2010, Chen_Gu_Wen_PRB_2010, Qi_RMP2011,  spt_chen_gu_wen2011,spt_schuch_garcia_cirac,spt_chen_gu_wen,spt_chen_gu_liu_wen,spt_chen_gu_liu_wen2013}
Phase transitions between distinct topological phases cannot be characterized by means of local order parameters,
whereas the structure of quantum entanglement serves as
a fingerprint for the phases.~\cite{li_haldane,spt_pollmann_oshikawa,  spt_pollmann_turner}
In particular, symmetry protected topological (SPT) phases are such phases that can be distinguished from trivial phases in the presence of certain symmetries.~\cite{spt_pollmann_oshikawa, spt_chen_gu_wen2011,spt_chen_gu_wen, spt_pollmann_turner}

In one-dimensional systems, the Haldane phase of
spin $S=1$ chains is a canonical example of SPT phases.\cite{HaldanePL1983,HaldanePRL1983,affleck1987,affleck1988}
Recently generalizations to SU($n$) symmetric systems have been developing.
According to the general classification theory based on group cohomology,  SU($n$) symmetric systems support $n-1$ distinct nontrivial $\mathbb{Z}_n$ SPT phases that are protected
by $Z_n \times Z_n$ symmetry (which is a subset of SU($n$) symmetry). \cite{spt_chen_gu_liu_wen2013,Duivenvoorden2013}

Certain cold fermionic
atoms, such as the alkaline-earth atoms and ytterbium
atoms, have their electronic degrees of freedom decoupled from their
nuclear spins 
and thus offer an ideal platform for realizing SU($n$) symmetric systems.\cite{fukuhara2007,cazalilla2009,desalvo2010,taie2010,Pagano2014,cazalilla2014,spt_sun_capponi_AP2016}
It has been shown that cold fermionic atoms loaded into an optical lattice show a Mott insulating phase with SU($n$) spin symmetry.\cite{gorshkov,taie2012}
The possible two electronic states of atoms, $^1S_0$ or $^3P_0$,
further lead to another internal degree of freedom, {\it orbital}.\cite{gorshkov,Zhang2014}
Stable isotopes of these atoms with half-odd-integer nuclear spin
can host a two-orbital
SU($n$) fermion system and also an SU($n$) spin system in an optical lattice.~\cite{gorshkov,Scazza2014}

One simple setup for the $\mathbb{Z}_n$ SPT phases is given by an SU($n$) spin chain with the local  ($n^2-1$)-dimensional bases
of adjoint representation,
which are made of the fundamental representation $\bm n$ and its conjugate $\bar{\bm n}$.
(When $n=2$, this phase corresponds to the $S=1$ Haldane phase.\cite{HaldanePL1983,HaldanePRL1983,affleck1987,affleck1988})
On these bases, a gapped valence-bond-solid (VBS) ground state appears in a wide parameter range of SU($n$) symmetric models
including  the exactly soluble SU$(n)$ Affleck-Kennedy-Lieb-Tasaki (AKLT) model.\cite{affleck1988,greiter2007a,greiter2007b,Katsura_JPA2008,morimoto_ueda_momoi_furusaki}
The ground state has  a non-local  string order\cite{string_order_den_nijs, string_kennedy_tasaki,Duivenvoorden2012,spt_znxzn_string_DuivenvoordenPRB2013,spt_znxzn_DuivenvoordenPRB2013}
and is well described by the finitely correlated state or matrix-product (MP) state.~\cite{fannes1992,mps_ostlund,mps_rommer}
This phase is protected by the $Z_n \times Z_n$ symmetry\cite{morimoto_ueda_momoi_furusaki} and characterized by the topological index $\pm 1\in \mathbb{Z}_n$.
In this phase, 
the bond-center inversion symmetry is broken, and the number of edge states is $n$.
Another class $\pm 2 \in \mathbb{Z}_n$ of $\mathbb{Z}_n$ SPT phases was also found to appear in the SU($n$) spin
systems in the adjoint
representation.\cite{roy2018}

The $(n^2-1)$-dimensional basis can be easily realized in the two-orbital SU($n$) fermion system.
This is achieved by setting the ground-state orbital density $n_g$ and
the excited-state orbital density $n_e$ to $n_g=n-1$ and $n_e=1$,
and then using decomposition for the local basis as
${\bm n} \otimes \bar{\bm n}=({\bm n^2-1}) \oplus {\bf 1}$.
However, the nature of the ground state of the corresponding low-energy effective model has not been fully explored yet.

In this paper, we present a systematic study of various SU(4) models defined on the bases
in the adjoint representation ${\bf 15}$, which arises from
${\bm 4}\otimes \bar{\bm 4} = {\bf 15}\oplus {\bf 1}$.
We first study the low-energy effective spin model of the two-orbital SU(4) Hubbard model
at filling $n_g=3$ and $n_e=1$,
which we call SU(4) $e$-$g$ spin model,
and compare with the $\mathbb{Z}_4$ SPT phase of the SU(4) AKLT model based on ${\bf 15}$ states.
Since local bases are the same ${\bf 15}$ representations,
one may naively expect that the ground state of the SU(4) $e$-$g$ spin model has the same topological
character as the SU(4) AKLT state, i.e., the ground state of the SU(4) AKLT
model,\cite{affleck1988,Katsura_JPA2008,morimoto_ueda_momoi_furusaki} which has the topological
index $\pm 1\in \mathbb{Z}_4$.
However, our numerical calculation shows that the ground state of  the SU($4$) $e$-$g$ spin model
belongs to the $\mathbb{Z}_4$ SPT phase
with the index $2\in \mathbb{Z}_4$.
Such an SPT phase is known to appear in different types of SU(4) cold atom systems which are defined on ${\bf 20}$ representation and are realized with the fillings  $n_g=2$ and $n_e=2$.~\cite{su4_nonne_totsuka,su4_bois_totsuka,spt_sun_capponi_AP2016}

To understand the origin of this topological character, we further study various SU(4) symmetric models by extending the parameter space,
which contains an SU(4) bilinear model [defined by a bilinear form of SU(4) generators]
and the SU(4) AKLT model.
We numerically study ground-state properties using the infinite-size density matrix renormalization group (iDMRG).~\cite{dmrg_white_a,dmrg_white_b,idmrg_ian}
In the obtained phase diagram,
we find two  $\mathbb{Z}_4$ SPT phases; one has the index $\pm 1 \in \mathbb{Z}_4$ and the other $2\in \mathbb{Z}_4$.
A topological phase transition appears
between these two phases, e.g.,
when we smoothly change the model
from the SU(4) $e$-$g$ spin model to the SU(4)
bilinear model.
The nature of this topological phase transition is further explored by studying the SU(4) bilinear-biquadratic model.
We find that the change of local interactions controls the antisymmetrization effect on bonds, and causes a change of edge states
at the topological transition point.
We quantify this antisymmetrization effect by using Casimir operators defined for neighboring spins.
In addition, we analytically construct the wave function of the SPT state with the index $2\in \mathbb{Z}_4$ by using MP representation.
This demonstrates that the SPT state with $2\in \mathbb{Z}_4$ has sixfold degeneracy at the edge and can be realized on the local bases made of $n_g=3$ and $n_e=1$ fermions, i.e., the representations $\bar{\bf 4}$ and ${\bf 4}$.

The paper is organized as follows: In Secs.~\ref{subsec:su_n_H}--\ref{subsec:SU(4)_e-g_hb_model},
we study  topological properties of  the SU($n$) $e$-$g$ spin model  and bilinear model.
In Sec.~\ref{sec:su_n_blbq}, we study the topological phase transition in the SU(4) bilinear-biquadratic model, controlling
the expectation value of the quadratic bond Casimir operator.
The ground state in the $\mathbb{Z}_4$ SPT phase with the index $2 \in \mathbb{Z}_4$ in the 15-dimensional adjoint representation is presented
in Sec.~\ref{sec:SU4MPS}.
We summarize this paper in Sec.~\ref{sec:summary}, showing a phase diagram of two topological phases.

\section{Topological characters of the SU($n$) $e$-$g$ spin model, bilinear model, and bilinear-biquadratic model}
\label{sec:model}

\subsection{SU($n$) $e$-$g$ spin model: Hamiltonian}
\label{subsec:su_n_H}
Some of alkaline-earth atoms, e.g., Ba, and an alkaline-earth-like atom Yb have stable isotopes with nuclear spin $I$ of half odd integers.
The electronic structure of these atoms has the singlet ground state $^1S_0$, where the electronic
angular momentum takes $J=0$. These atoms also have a metastable excited
state in a singlet $^3P_0$ ($J=0$). Thus, nuclear spins are decoupled from the electronic states
and these atoms can be regarded as SU($n$) fermions with two orbitals,
where $n$ corresponds to the number of nuclear spin states, e.g., $n=2I+1$.
The ground state is called $g$-orbital and the excited state $e$-orbital.
Aligning these atoms in an optical lattice realizes a chain of two-orbital SU($n$) fermions.\cite{gorshkov} By loading atoms with
only selected
species of nuclear spins, we can set $n$ to be an arbitrary number in $1\le n \le 2I+1$.\cite{Pagano2014}

\begin{figure}[t]
\includegraphics[width=8.cm]{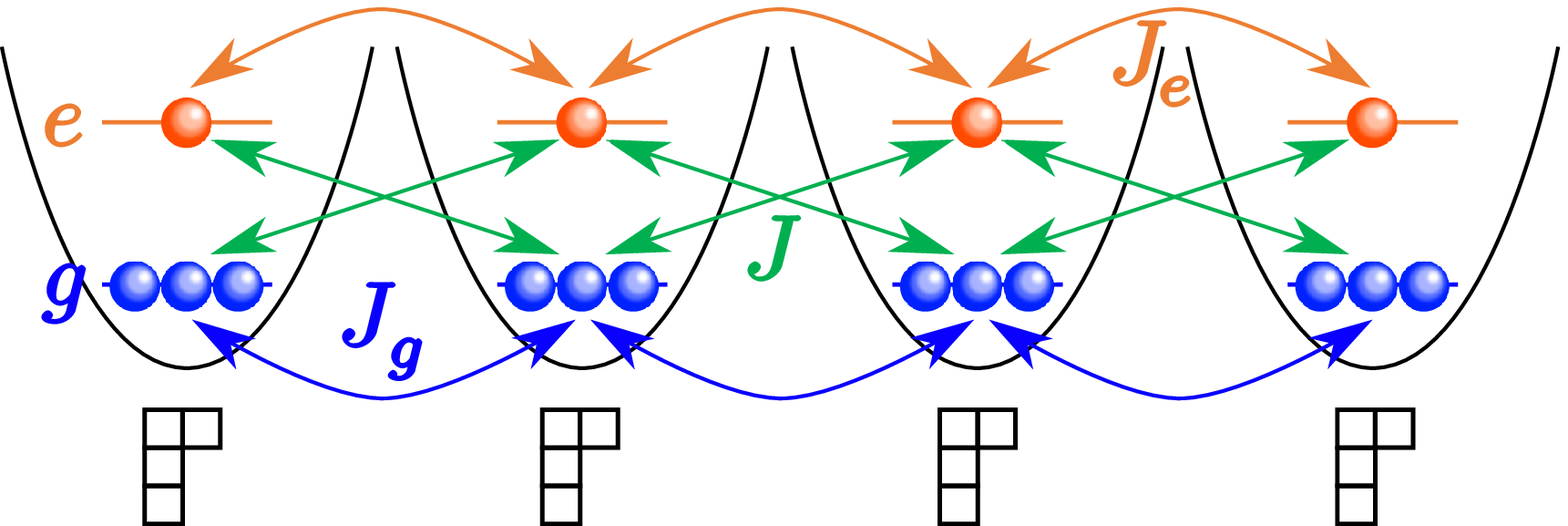}
\caption{(Color online)
Schematic picture of SU(4) symmetric models at filling $n_g=3$ and $n_e=1$.
Hund's rule coupling selects the 15-dimensional representation as the low-energy local basis in each site.
$J_\alpha$ ($\alpha=g,e$) and $J$ denote the intraorbital and
interorbital exchanges, respectively.
Setting $J=0$ realizes the SU(4) $e$-$g$ spin model given in
Eq.~(\ref{eq:H_eg}), whereas setting $J_e=J_g=J$ realizes the SU(4) bilinear model given in Eq.~(\ref{eq:H_SU4}).
}
\label{fig:e-g_model}
\end{figure}
At the filling density of $g$ orbital $n_g=n-1$ and of $e$ orbital
$n_e=1$, local single-site states are
in the fundamental representation ${\bm n}$ in $e$ orbital and its conjugate $\bar{\bm n}$
in $g$ orbital.
The total single-site basis can be decomposed into two irreducible representations as
$\bar{\bm n}\otimes{\bm n}=({\bm n}{\bf {}^2-1})\oplus {\bf 1}$.
In the presence of Hund's rule coupling,
the $(n^2-1)$-dimensional representation is selected as the
single-site basis in low-energy regime. (See Fig.~\ref{fig:e-g_model}.)
In the Mott insulator, the low-energy effective Hamiltonian
defined on these $(n^2-1)$-dimensional bases is derived with
the second-order perturbational expansion as
\begin{align}
H_{\mbox{\it e-g}}= \sum_{\alpha=g,e}J_\alpha \sum_{j}\sum_{m,m^\prime=1}^n
S^{m^\prime}_m (j,\alpha) S^m_{m^\prime} (j+1,\alpha)
\label{eq:H_eg}
\end{align}
with $J_\alpha>0$, where
\begin{equation}
S^m_{m^\prime}(j,\alpha)=c_{j\alpha m^\prime}^\dagger c_{j\alpha m}
\end{equation}
($m,m^\prime=1,\cdots,n$) denote the operators of SU($n$) algebra on $\alpha$-orbital states
with
the annihilation operator $c_{j\alpha m}$ of a fermion on $j$th site in $\alpha$ orbital
with nuclear spin index $m$.
We call Eq.~(\ref{eq:H_eg}) ``SU($n$) $e$-$g$ spin model"  in this paper.
We show a schematic picture of this model with $n=4$ in Fig.~\ref{fig:e-g_model}.

\subsection{SU($n$) $e$-$g$ spin model: Entanglement spectrum and topological indexes}
\label{subsec:Entanglement spectrum and topological index}

We study the topological properties of the ground state of SU($n$) $e$-$g$ spin model with $n=3$ and $4$,
by using the entanglement spectrum  and
also the topological indices associated with $Z_n \times Z_n$ symmetry actions and
inversion.
The entanglement spectrum and topological indices are obtained by iDMRG calculations.~\cite{li_haldane,spt_pollmann_turner}
The entanglement spectrum effectively corresponds to the energy spectrum for edge states. In particular, its degeneracy corresponds to the existence of nontrivial edge states that characterize the SPT phases.~\cite{Kennedy_1990,Miyashita_Yamamoto_PRB1993,White_Huse_PRB1993,Chen_Gu_Wen_PRB_2010,spt_chen_gu_liu_wen,spt_sun_capponi_AP2016}
We first describe the results of the SU(3) symmetric case, briefly explaining
our method for the identification of SPT phases.
After that we show our numerical results of the SU(4) symmetric case.

\subsubsection{SU(3) symmetric case}

The entanglement spectrum is easily obtained through the iDMRG calculations, because the fixed point of iDMRG gives us a (variational) ground state as a form of Schmidt decomposition
\begin{equation}
\ket{\Psi} = \sum_i \lambda^{~}_{i} \ket{\psi^{~}_{i}} \ket{\phi^{~}_{i}}
\end{equation}
with the Schmidt coefficients $\lambda^{~}_{i}$ and the orthonormal basis $\ket{\psi^{~}_{i}}$ ($\ket{\phi^{~}_{i}}$) for the left (right) semi-infinite system.
The sequence of the ``energy levels"
$\zeta^{~}_i$,
defined by
\begin{equation}
\zeta^{~}_i = -2 \log \lambda^{~}_i,
\label{eq:e_spectrum}
\end{equation}
gives the entanglement spectrum.~\cite{li_haldane}
The number of degeneracy for each level is related
to the dimension of the corresponding irreducible
representation of SU($n$) symmetry group. The dimensions for both SU(3) and SU(4)
symmetry groups are listed in Appendix~\ref{appendix:dimensions}.

In a topological phase,
the lowest level has a certain degeneracy which corresponds to the number of edge states.
A limited number of irreducible representations appear in a lower part of the entanglement
spectrum. This characteristic sequence serves as a fingerprint of each topological
phase.\cite{li_haldane,spt_pollmann_oshikawa}
It was numerically found~\cite{morimoto_ueda_momoi_furusaki} that, when the ground state is an SPT state protected by $Z_3 \times Z_3$ symmetry, all the numbers of degeneracy are multiples of three
and the lowest level has threefold degeneracy, which comes from the number of edge states,
i.e.,
the dimension of the representation ${\bf 3}$ or $\bar{\bf 3}$.
It was also proved\cite{spt_pollmann_oshikawa} that when the ground state has the bond-center inversion symmetry with the odd parity,
all the numbers of degeneracy are multiples of two.

\begin{figure}
\includegraphics[width=7cm]{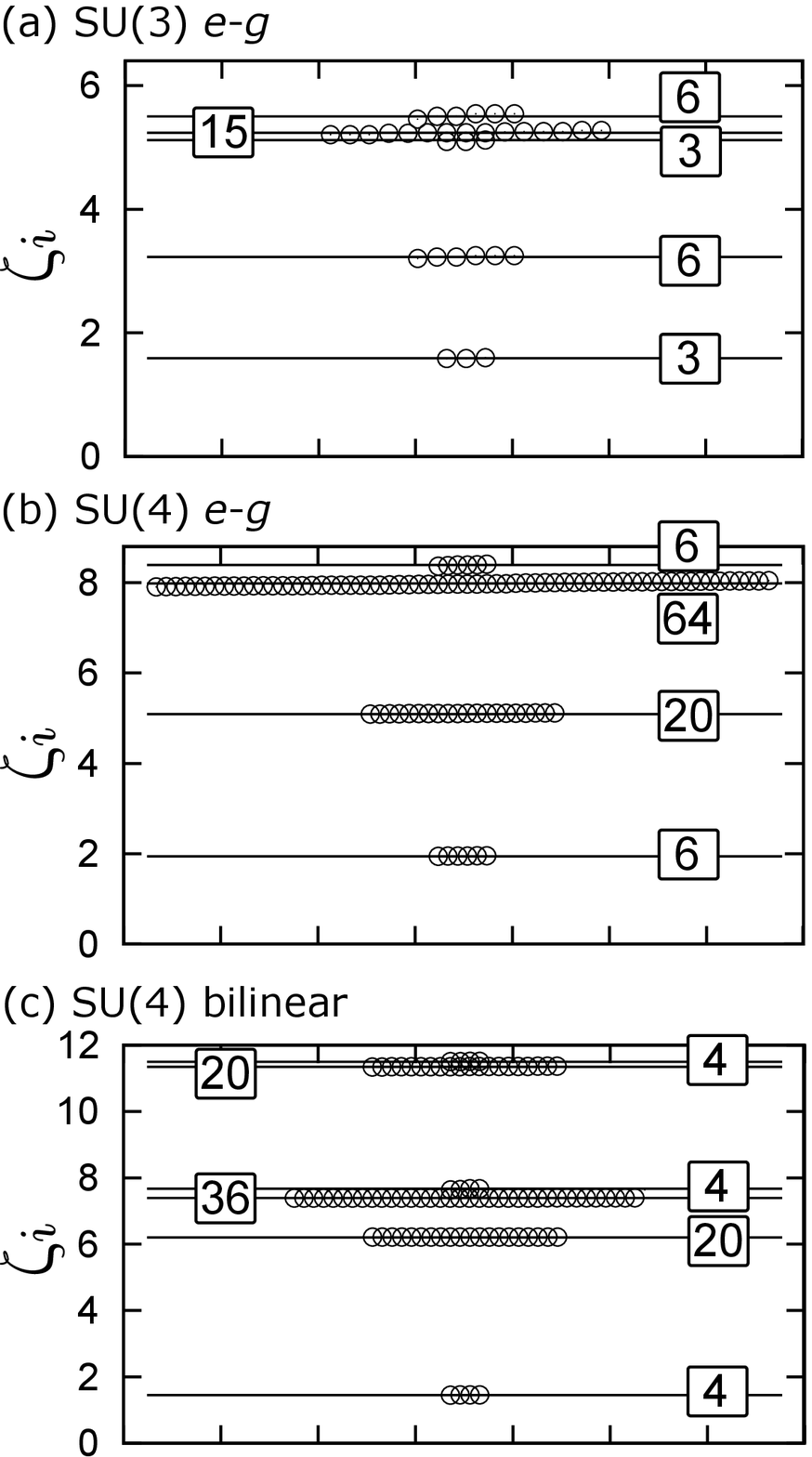}
\caption{Entanglement spectra in (a) the SU(3) $e$-$g$ spin model, (b) the SU(4) $e$-$g$ spin model,
and (c) the SU(4) bilinear model.
The numbers enclosed in squares denote the degeneracy factor of each level.
The dimension $\chi$ of the auxiliary (edge state) space is chosen as (a) $300$, (b)$225$,
and (c) 225. }
\label{fig:e_spec}
\end{figure}

Figure \ref{fig:e_spec}(a) shows our result of the entanglement spectrum for the ground state of the SU(3) $e$-$g$ spin model
in the case $J_e/J_g=1$.
The lowest level has threefold degeneracy, which means the number of edge states is three.
All the levels in the spectrum have the degeneracies of multiples of three, e.g.,
3, 6, and 15, which can be found in the list of dimensions $D^{nm}$ shown in Table~\ref{table:dim_irrep} in Appendix~\ref{appendix:dimensions}.
In addition, we find that there are doubly degenerate ground states that are related by inversion operation with each other.
All of these properties are
the same as those\cite{morimoto_ueda_momoi_furusaki} found in the $\mathbb{Z}_3$ SPT phase
with the index $\pm 1\in \mathbb{Z}_3$.

Next we study
the factor system of two symmetry operations of the group $Z_n \times Z_n$ [which is a subset of SU(n) symmetry]
in the projective space, to identify
the topological index.
Here, we briefly describe how to obtain the index. Let us consider the symmetry actions of $Z_n \times Z_n$
performed by the generators $x$ and $y$ of the two $Z_n$ groups.
These satisfy the relations $x^n=y^n=1$ and $[x,y]=0$. In the MP
representation, finitely correlated ground states in
an $L$-site system under the periodic boundary condition can be written as
\begin{equation}
\ket{\Psi} = \sum_{\{ \sigma_j \}} \Tr \prod_j \big(
A^{\sigma_j} \big) \ket{\sigma_1 \sigma_2 \cdots \sigma_L},
\end{equation}
where $A^{\sigma_j}$ denote $\chi \times \chi$ matrices
specified by the local states $\sigma_j$ on $j$th site.
The symmetry action  $\hat{g}=x$, $y$ transforms the matrix as
\begin{equation}
\hat{g} \ket{\Psi} = \sum_{\{ \sigma_j \}} \Tr \prod_j \Big( \sum_{\sigma'_j} g_{\sigma_j \sigma'_j} A^{\sigma'_j} \Big) \ket{\sigma_1 \sigma_2 \cdots \sigma_L}.
\end{equation}
When $\ket{\Psi}$ is invariant under the symmetry action of $\hat{g}$, i.e., $|\langle \hat{g} \rangle| := | \bra{\Psi}\hat{g} \ket{\Psi} | = 1$,
the transformation law in the projective representation is given by\cite{garcia_wolf_sanz_verstraete_cirac}
\begin{equation}
\sum_{\sigma'_j} g_{\sigma_j \sigma'_j} A^{\sigma'_j} = e^{i \theta_g} U^{-1}_g A^{\sigma_j} U^{~}_g
\end{equation}
with a discrete phase $\theta_g = 2\pi \ell /n$ ($\ell=0,\cdots,n-1$) and
a $\chi \times \chi$ matrix $U_g$. The  matrix $U_g$ performs  the symmetry action $\hat{g}$ to the matrix
$A^{\sigma_j}$ in the projective representation.
A factor system of this representation is given by
\begin{equation}
U_x U_y = e^{i\phi_g} U_y U_x
\end{equation}
with a discrete phase $\phi_g = 2\pi \mathcal{O}_{Z_n \times Z_n}/n$. Here
$\mathcal{O}_{Z_n \times Z_n}$ is an integer defined in the range
\begin{equation}
\mathcal{O}_{Z_n \times Z_n}=\lfloor -n/2 \rfloor+1,\cdots, \lfloor n/2 \rfloor,
\label{eq:ell}
\end{equation}
where $ \lfloor \cdots \rfloor$ denotes the floor function.
When the index $\mathcal{O}_{Z_n \times Z_n}$ has a non-zero value, the ground state is topologically
nontrivial and this index classifies its phase into the $\mathbb{Z}_n$ SPT phase with the topological index
$\mathcal{O}_{Z_n \times Z_n} \in \mathbb{Z}_n$.
We can obtain the topological index by
\begin{equation}
\mathcal{O}_{Z_n \times Z_n} = \frac{n}{2 \pi i } \log \left[\frac{1}{\chi} \Tr \left( U_xU_yU^{-1}_xU^{-1}_y \right) \right]~.
\end{equation}
The numerical method to calculate the matrix $U_g$ is detailed in Ref.~[\onlinecite{spt_pollmann_turner}].
We note that the classification of the topological phases from the factor systems of the projective representation is equivalent to the classification using conventional nonlocal string orders.~\cite{spt_pollmann_oshikawa, spt_pollmann_turner, Duivenvoorden2012}

\begin{figure}
\includegraphics[width=8.5cm]{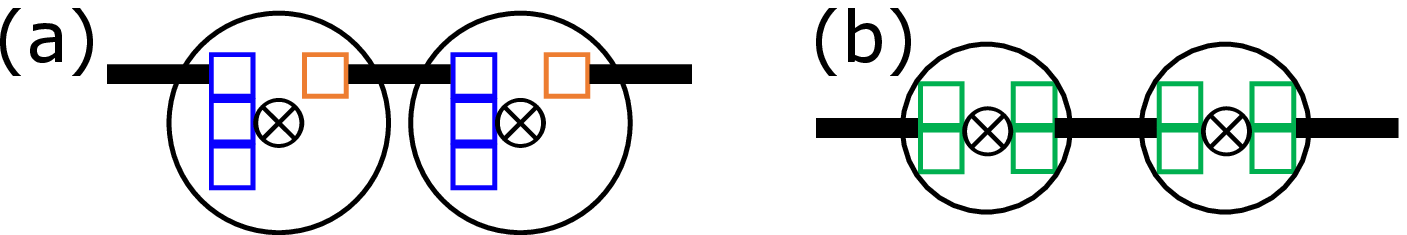}
\caption{(Color online) Schematic figures of two possible valence bond solid (VBS) states with the topological index
$\pm1 \in \mathbb{Z}_4$ (a) and with $\pm2 \in \mathbb{Z}_4$ (b)
on the local bases in the 15-dimensional representation.
The black circle on each site denotes the projector onto the irreducible representation {\bf 15}.
In (a),  each site basis is given by the product of the fundamental representation ${\bf 4}$ of $e$ orbital
and its conjugate $\bar{{\bf 4}}$ of $g$ orbital,  and, in (b), by the product of  the two representations
${\bf 6}$, which are formed by the two (antisymmetric) fermion pairs out of $n_g=3$ and $n_e=1$ fermions.
Each thick horizontal bond connecting the two neighboring diagrams forms the singlet.}
\label{fig:aklt_states}
\end{figure}

We numerically evaluate the value of $\mathcal{O}_{Z_3 \times Z_3}$ in the ground state of the SU(3) $e$-$g$ spin model using the iDMRG method
and obtain
\begin{equation}
\mathcal{O}_{Z_3 \times Z_3} = \pm 1.
\end{equation}
We confirm that the ground states are doubly degenerate and these two can be exchanged with each other by acting
the inversion operation.
All of these results show that the ground state is in the $\mathbb{Z}_3$ SPT phase with the index $\pm 1 \in \mathbb{Z}_3$ and
is topologically identical with the ground state of the SU(3) bilinear mode studied in Ref.~[\onlinecite{morimoto_ueda_momoi_furusaki}].
A schematic picture of the MP state with $\pm1 \in \mathbb{Z}_n$ (for the case of $n=4$) is
shown in Fig.~\ref{fig:aklt_states}(a).
With varying the couplings $J_e$ and $J_g$, we further confirm that  these SPT characters do not change in the range of $0.5 \leq J_e/J_g \leq 999$.

\subsubsection{SU(4) symmetric case}
Now we study the ground state of the SU(4) $e$-$g$ spin model.
Figure \ref{fig:e_spec}(b) shows the entanglement spectrum for the case $J_e/J_g=1$.
We find that the low-lying levels do not always have the degeneracy of multiples of four
and the lowest level has sixfold degeneracy, which suggests that the number of edge states is six.
This contrasts to the fourfold degeneracy (corresponding to ${\bf 4}$ or $\bar{\bf 4}$)
in the $\mathbb{Z}_4$ SPT phase with $\pm 1 \in \mathbb{Z}_4$ that was studied in
Ref.~[\onlinecite{morimoto_ueda_momoi_furusaki}].
The numbers of degeneracy are labeled in the figure, which are related to the dimensions of the irreducible representations
of the SU(4) group shown in Appendix~\ref{appendix:dimensions}.
Furthermore, the ground state has the bond-center inversion symmetry.
All of these results are
inconsistent with the $\mathbb{Z}_4$ SPT phase with the index $\pm 1 \in \mathbb{Z}_4$.

\begin{figure}
\includegraphics[width=8.5cm]{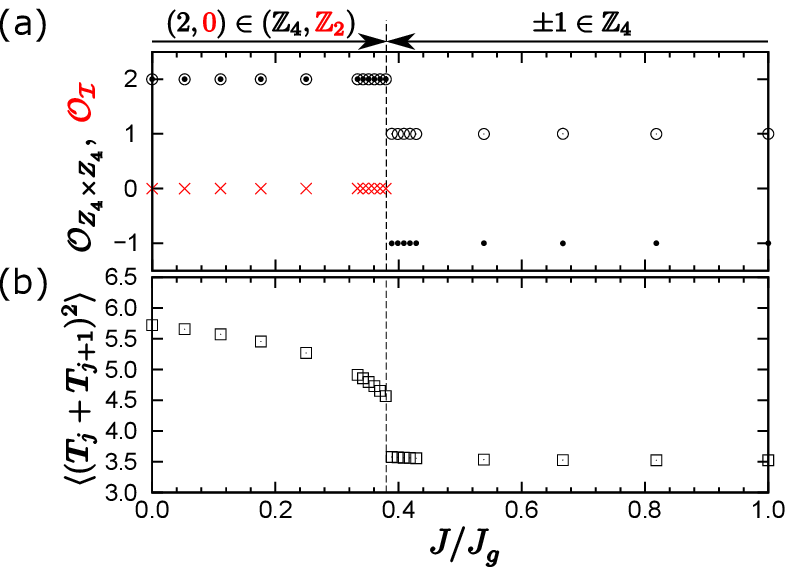}
\caption{(Color online) (a) Topological indices $\mathcal{O}_{Z^{~}_{4} \times Z^{~}_{4}}$ and $\mathcal{O}_\mathcal{I}$, and
(b) expectation values of the quadratic bond Casimir operator
$\langle ({\bm T}_j +{\bm T}_{j+1})^2 \rangle$
as a function of
the coupling $J/J^{~}_{\rm g}$ given in the extended SU(4) Hamiltonian~(\ref{eq:hamiltonian_eg_hb}).
The case $J=0$ corresponds to the SU(4) $e$-$g$ spin model and $J/J_g=1$ to the SU(4) bilinear model.
The symbols $\bigcirc$, $\bullet$, and \textcolor{red}{$\times$} represent the indexes
$\mathcal{O}_{Z^{~}_{4} \times Z^{~}_{4}}$ of $A^{\sigma}$, $\mathcal{O}_{Z^{~}_{4} \times Z^{~}_{4}}$ of $(A^{\sigma})^{T}$, and $\mathcal{O}_\mathcal{I}$, respectively.
The auxiliary space of iDMRG calculations is set as $\chi=225$.
}
\label{fig:phase_diagram_su4eg}
\end{figure}

To elucidate the topological properties of the ground state, we numerically evaluate the index $\mathcal{O}_{Z_4 \times Z_4}$ and obtain
\begin{equation}
\mathcal{O}_{Z_4 \times Z_4} = 2,
\end{equation}
as shown  in Fig.~\ref{fig:phase_diagram_su4eg}(a).
We thus find that the ground state is in the $\mathbb{Z}_4$ SPT phase with the index $2\in \mathbb{Z}_4$.
VBS ground states in this topological class ($2\in \mathbb{Z}_4$) were previously studied in SU(4) models
on the local bases of the 20-dimensional representation~\cite{su4_nonne_totsuka, su4_bois_totsuka} and 15-dimensional
representation,\cite{roy2018}
where edge states are in the representation ${\bf 6}$ or $\bar{\bf 6}$ and sixfold degenerate.
This degeneracy is also consistent with our analysis of the entanglement spectrum shown in Fig.~\ref{fig:e_spec}(b).
In Sec.~\ref{sec:SU4MPS},
we will explicitly construct MP states that show sixfold degenerate edge states arising from the three $g$-orbital fermions and one $e$-orbital fermion on each site and
form a bond-inversion symmetric state.

In addition, since the ground state has bond-center inversion symmetry,
we investigate the factor system of the inversion, i.e., an index $\mathcal{O}_{\mathcal{I}}$ (for details, see Appendix \ref{app:inversion}).
We numerically evaluate $\mathcal{O}_{\mathcal{I}}$ for the ground state of the SU(4) $e$-$g$ spin model and obtain the estimate $\mathcal{O}_{\mathcal{I}}=0$,
as shown in Fig.~\ref{fig:phase_diagram_su4eg}(a).
This indicates that this phase belongs to a trivial phase in terms of the inversion symmetry.

We thus find that the ground state of the SU(4) $e$-$g$ spin model is in the $\mathbb{Z}_4$ SPT phase with the index $2 \in \mathbb{Z}_4$ and inversion symmetric with $\mathcal{O}_{\mathcal{I}}=0$.
These topological properties are the same as those observed in the VBS ground state of
an SU(4) model
on the local bases of the 20-dimensional representation.\cite{su4_nonne_totsuka, su4_bois_totsuka}
We have further confirmed that these SPT characters remain in the range
$0.1 \leq J_e/J_g \leq 10$.

\subsection{SU($n$) bilinear model}
\label{subsec:su(n) bilinear model}

From the above calculations, we find that
the ground state of the SU(4) $e$-$g$ spin model is in the $\mathbb{Z}_4$  SPT phase with the topological index
$2 \in \mathbb{Z}_4$, whereas that of the SU(3) model is in the SPT phase with
$\pm 1 \in \mathbb{Z}_3$.
To understand why topological characters are different between SU(3) and SU(4) $e$-$g$ spin models, we study topological properties of
the SU(3) and SU(4) bilinear models, respectively, on the bases of eight-dimensional and 15-dimensional representations as references.

The minimal SU($n$) symmetric Hamiltonian has a bilinear form of SU($n$) algebra given by
\begin{align}
H=  J \sum_{j}\sum_{m,m^\prime=1}^n
S^{m^\prime}_m (j) S^m_{m^\prime} (j+1),
\label{eq:H_SU4}
\end{align}
where $J>0$ and
\begin{align}
S^m_{m^\prime} (j)=S^m_{m^\prime} (j,g)+S^m_{m^\prime} (j,e)
\end{align}
are defined on the bases of ($n^2-1$)-dimensional representations.
We can rewrite the Hamiltonian~(\ref{eq:H_SU4}) as
\begin{align}
H=  2 J \sum_{j} {\bm T}_j\cdot {\bm T}_{j+1}
\end{align}
using the SU($n$) generators
$T_j^\gamma$ ($\gamma=1,\cdots,n^2-1$)
in the ($n^2-1$)-dimensional adjoint representations.
The local interaction corresponds to the quadratic Casimir operator $({\bm T}_j + {\bm T}_{j+1})^2$ on bond ($j,j+1$).
To convert the $e$-$g$ spin model to the above SU($n$) bilinear model, one needs to impose $J_e=J_g(=J)$ and also include the same amount of exchange terms between
different orbitals,
\begin{equation}
J \sum_{m,{m^\prime}} \{S^{m^\prime}_m (j,e) S^m_{m^\prime} (j+1,g)
+ S^{m^\prime}_m (j,g) S^m_{m^\prime} (j+1,e)\},
\label{eq:J_eg}
\end{equation}
as shown in Fig.~\ref{fig:e-g_model}.

The topological property of the SU(3) bilinear model has been studied in Ref.~[\onlinecite{morimoto_ueda_momoi_furusaki}];
the ground state is in the topological phase
protected by $Z_3 \times Z_3$ symmetry and it has the topological index $\pm1 \in \mathbb{Z}_3$. The ground states are doubly degenerate
and do not have bond-center inversion symmetry.

In the SU(4) bilinear model, we show the entanglement spectrum in Fig.~\ref{fig:e_spec}(c).
The lowest level has the fourfold degeneracy and all levels in the spectrum have the degeneracy
of multiples of four. This means that the number of edge states is four.
We numerically evaluate the index $\mathcal{O}_{Z_4 \times Z_4}$ and obtain the estimate
$\mathcal{O}_{Z_4 \times Z_4} = \pm 1$ as shown
in Fig.~\ref{fig:phase_diagram_su4eg}(a).
Also, we find that the ground states are doubly degenerate, which  breaks the bond-center inversion symmetry, and these two
can be exchanged with each other by the inversion operation.
These results conclude that the ground state of the SU(4) bilinear Hamiltonian is in the $\mathbb{Z}_4$ SPT phase with the index $\pm1 \in \mathbb{Z}_4$, which is topologically distinct from the ground state of the SU(4) $e$-$g$ spin model, but identical with that of
the SU(4) AKLT model.\cite{Katsura_JPA2008,morimoto_ueda_momoi_furusaki}

\subsection{From SU(4) $e$-$g$ to  bilinear model: Topological phase transition}
\label{subsec:SU(4)_e-g_hb_model}

From the analyses in Secs.~\ref{subsec:Entanglement spectrum and topological index} and \ref{subsec:su(n) bilinear model},
we find that the ground states of the SU(4) $e$-$g$ spin model and the SU(4) bilinear model are topologically distinct from each other.
To see why and how topological characters change, we deform the Hamiltonian from the SU(4) $e$-$g$ spin model to the
SU(4) bilinear model continuously and study a topological phase transition.

We consider the following extended Hamiltonian:
\begin{equation}
\mathcal{H} = H_{e\mathchar`-g} + J \sum_{\alpha \neq \beta} \sum_{j,m,n} S^{n}_m(j,\alpha) S^{m}_{n}(j+1,\beta),
\label{eq:hamiltonian_eg_hb}
\end{equation}
setting the couplings $J_g=J_e$ in the Hamiltonian $H_{e\mathchar`-g}$ [Eq.~(\ref{eq:H_eg})].
This Hamiltonian ${\cal H}$ includes the SU(4) $e$-$g$ spin model  at $J=0$ and the SU(4) bilinear model
at $J=J_g$.
We numerically investigate the ground state of $\mathcal{H}$, Eq.~(\ref{eq:hamiltonian_eg_hb}), and
obtain the indexes  $\mathcal{O}_{Z_4 \times Z_4}$ and  $\mathcal{O}_{\cal I}$ as a function of $J/J_g$.
The results are
shown in Fig.~\ref{fig:phase_diagram_su4eg}(a);
there occurs a clear phase transition around $J/J_g=0.38$, where both the breaking of
bond-center inversion symmetry and the discontinuous transition
of the topological index of the $Z_4 \times Z_4$ symmetry occur, simultaneously.

Then what controls the topological phase transition?
To see the mechanism of this transition, we focus on the value of the quadratic Casimir operators
$\langle ( {\bm T}_j + {\bm T}_{j+1} )^2 \rangle$ on bonds.
Our numerical results are shown in Fig.~\ref{fig:phase_diagram_su4eg}(b).
With increasing $J$, the values of the Casimir operator gradually decrease and show a drastic
change, at the transition point, from a high ``spin" phase, where the Casimir operator has
a large expectation value, to a low spin phase, where it has a smaller value.
The insertion of the interorbital coupling $J$ thus enhances antisymmetrization in bond degrees of freedom, and
thereby induces this phase transition from
a high spin SPT phase to a low spin SPT phase.

Our numerical results show that the $\mathbb{Z}_4$ SPT phase with the index $2 \in \mathbb{Z}_4$ takes a larger value
of the Casimir operator than the  $\mathbb{Z}_4$ SPT phase with the index $\pm 1 \in \mathbb{Z}_4$,
\begin{equation*}
\langle ( {\bm T}_j + {\bm T}_{j+1} )^2 \rangle_{\pm1\in \mathbb{Z}_4} <\langle ( {\bm T}_j + {\bm T}_{j+1} )^2 \rangle_{2 \in \mathbb{Z}_4}.
\end{equation*}
We can also see the same tendency in the SU(4) AKLT  state of $\pm1 \in \mathbb{Z}_4$~\cite{Katsura_JPA2008, morimoto_ueda_momoi_furusaki}
and that of $2 \in \mathbb{Z}_4$  as shown
in Sec.~\ref{sec:SU4MPS_corr}.
This tendency can be understood by decomposing the bond degrees of freedom into irreducible
representations of the SU(4) group;
bond states of the SU(4) AKLT state with $\pm 1 \in \mathbb{Z}_4$ belong only to
the representations ${\bf 1}$ and
${\bf 15}$, whereas those of the exact VBS state with $2\in \mathbb{Z}_4$ belong to
${\bf 1}$, ${\bf 15}$, or ${\bf 20}$.
(See Fig.~\ref{fig:aklt_states} and also Sec.~\ref{sec:SU4MPS}.)
This discrepancy of the bond bases results in the difference of expectation values of the bond
Casimir operators.

The above results lead to a promising anticipation that controlling the value
$\langle ( {\bm T}_j + {\bm T}_{j+1} )^2 \rangle$ can induce the phase transition
between $\pm1 \in \mathbb{Z}_4$ phase and $2 \in \mathbb{Z}_4$ phase.

\subsection{SU($4$) bilinear-biquadratic model}
\label{sec:su_n_blbq}

To further examine our idea that the two topological phases have a distinguishable difference in
magnitude of the values of the Casimir operator
and any control of these values can induce topological phase transitions,
we study the SU(4) bilinear-biquadratic model, whose Hamiltonian reads
\begin{align}
H_{\rm BBQ}= 2J & \sum_{j}  \left[ \cos \theta
 ({\bm T}_j \cdot {\bm T}_{j+1})
+ \sin \theta \left( {\bm T}_j \cdot {\bm T}_{j+1} \right)^2
\right],
\label{eq:H_BBQ}
\end{align}
where $J>0$.

Changing the phase $\theta$ in the Hamiltonian~(\ref{eq:H_BBQ}),
we can control the expectation value of the quadratic Casimir operators on bonds.
This can be readily seen by rewriting $H_{\rm BBQ}$ in $0<\theta<\pi/2$ as
\begin{equation}
H_{\rm BBQ}=\frac{J}{2} \sin \theta \sum_j \Big[ \big( {\bm T}_j + {\bm T}_{j+1} \big)^2
- \lambda \Big]^2
\label{eq:bbq2}
\end{equation}
with $\tan\theta = 1/(8-\lambda)$ in $\lambda<8$, except for a constant term.
The minimum point of the quadratic function $f(x)=(x-\lambda)^2$ increases
with increasing $\lambda$ in $\lambda<8$ (or equivalently increasing $\theta$ in $0<\theta<\pi/2$).
We thus expect that the expectation value
$\langle ( {\bm T}_j + {\bm T}_{j+1} )^2 \rangle$ increases if we increase $\theta$ in $0<\theta<\pi/2$.

We already know that the ground state in a small positive $\theta$ regime belongs to the $\mathbb{Z}_4$ SPT
phase with the index $\pm 1 \in \mathbb{Z}_4$.
The SU(4) bilinear-biquadratic Hamiltonian at $\theta=0$ is obviously equivalent to the SU(4) bilinear Hamiltonian of  Eq.~(\ref{eq:H_SU4}),
whose ground state is in the $\mathbb{Z}_4$ SPT
phase with the index $\pm 1 \in \mathbb{Z}_4$, as shown in Sec.~\ref{sec:model}.
Also, the exact ground state of the SU($n$) bilinear-biquadratic Hamiltonian
is known to have a VBS wave function for general $n$ at
$\theta={\rm arctan}(2/3n)$.\cite{Rachel2010}  In the SU(4) symmetric case, this VBS ground state
at $\theta={\rm arctan}(2/6)\sim0.0525 \pi$ is again in the $\mathbb{Z}_4$ SPT
phase with the index $\pm 1 \in \mathbb{Z}_4$.\cite{morimoto_ueda_momoi_furusaki}

\begin{figure}
\includegraphics[width=8.5cm]{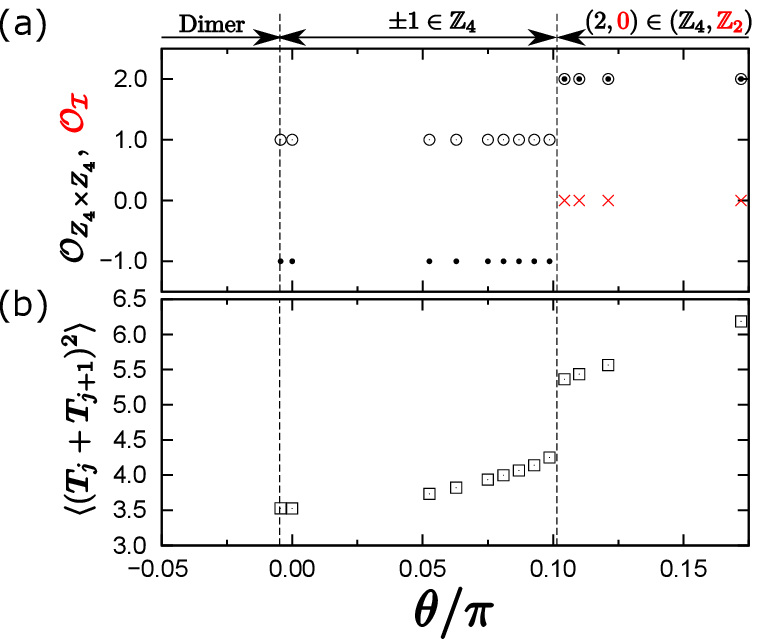}
\caption{(Color online) Phase diagram of the SU(4) bilinear-biquadratic model given in Eq.~(\ref{eq:H_BBQ}) with the
data of (a) indices $\mathcal{O}_{Z^{~}_{4} \times Z^{~}_{4}}$ and $\mathcal{O}_\mathcal{I}$,
and (b) expectation values of the quadratic bond-Casimir operator.
The symbols have the same meanings as those in Fig.~\ref{fig:phase_diagram_su4eg}.
The case $\theta=0$ corresponds to the SU(4) bilinear model and $\theta/\pi \sim 0.0525$ to the SU(4) AKLT
model whose exact ground state is a VBS state.\cite{Rachel2010,morimoto_ueda_momoi_furusaki}
The iDMRG calculation was performed with $\chi=100$.}
\label{fig:phase_diagram_blbq}
\end{figure}

We numerically study the ground state of the Hamiltonian $H_{\rm BBQ}$ varying $\theta$
around $\theta \approx 0$.
The $\theta$ dependence of the indices $\mathcal{O}_{Z^{~}_{4} \times Z^{~}_{4}}$ and $\mathcal{O}_{\mathcal{I}}$ is shown in Fig.~\ref{fig:phase_diagram_blbq}(a).
In our calculation of iDMRG up to $\chi=100$,
the index $\mathcal{O}_{Z^{~}_{4} \times Z^{~}_{4}}$ takes the values $\pm 1$
in the range of $-0.005 \leq \theta/\pi \leq 0.10$. Slightly above $\theta = 0.10\pi$, we again find the transition, where the recovering of
inversion symmetry and the discontinuous transition of the topological index $\mathcal{O}_{Z^{~}_{4} \times Z^{~}_{4}}$ with respect to
the $Z_4 \times Z_4$ symmetry occur simultaneously as same as found in Sec.~\ref{subsec:SU(4)_e-g_hb_model}.
The new phase that appears in $0.10 \lesssim \theta/\pi $ is topologically the same phase as found in the SU(4) $e$-$g$ spin model.
Thus the ground state is in the $\mathbb{Z}_4$ SPT phase with the index $\pm 1 \in \mathbb{Z}_4$ in the range $-0.005 \leq \theta/\pi \lesssim 0.10$ and in the
$\mathbb{Z}_4$ SPT phase with the index $2 \in \mathbb{Z}_4$ in the range $0.10 \lesssim \theta/\pi$ (we have confirmed that this phase extends to at least 0.17 as shown in Fig.~\ref{fig:phase_diagram_blbq}).
Moreover, as we expected, the expectation value
$\langle ( {\bm T}_j + {\bm T}_{j+1} )^2 \rangle$ monotonically increases with increasing $\theta$
as shown in Fig.~\ref{fig:phase_diagram_blbq}(b) and suddenly jumps (increases) at the
transition point.
Thus the increase of the expectation value of the Casimir operator induces the topological phase transition.

\begin{figure}
\includegraphics[width=8.5cm]{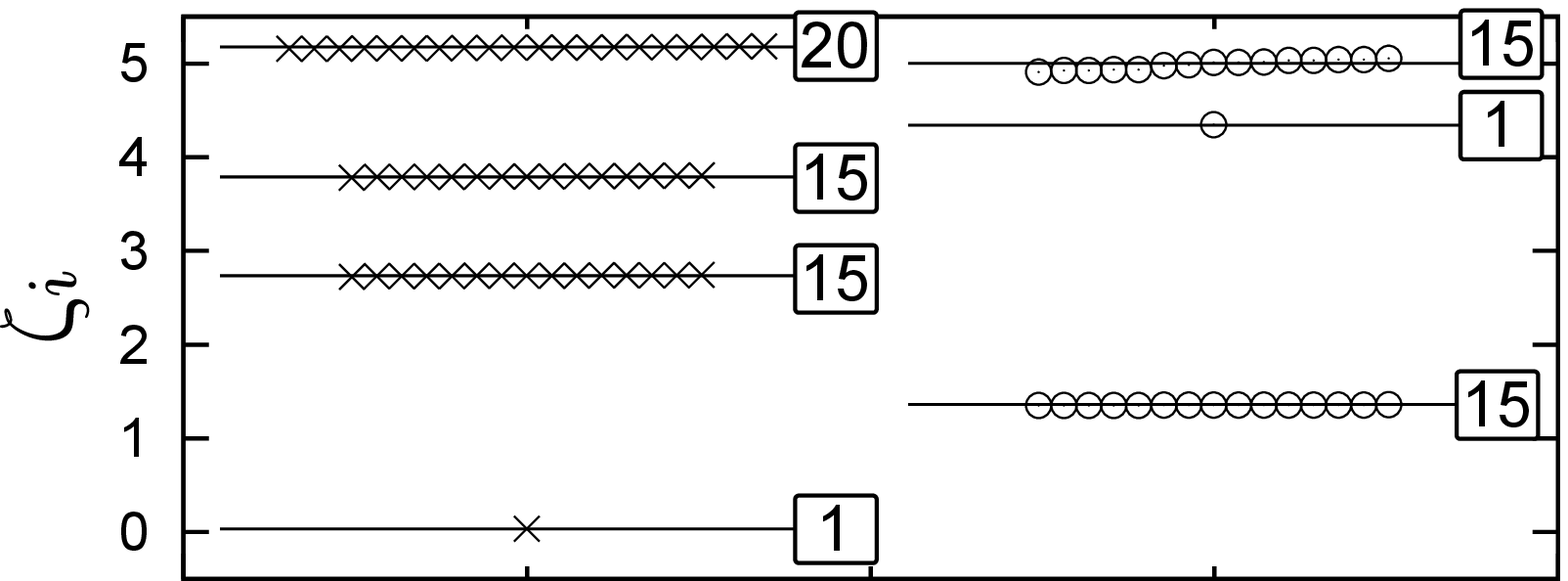}
\caption{Entanglement spectrum for the dimer phase at $\theta=-0.1\pi$ in the SU(4) bilinear-biquadratic model.
The left/right data show the spectrum of the chain divided without/with cutting a singlet dimer.
The iDMRG calculation was performed with $\chi=100$.
The slight deviation of the levels found in the top of the right panel is attributed to a finite-$\chi$ effect.}
\label{fig:spectrum_dimer}
\end{figure}

\begin{figure}
\includegraphics[width=8.5cm]{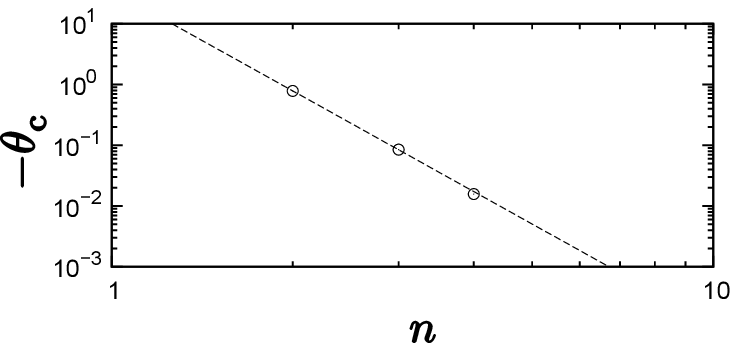}
\caption{Transition point $\theta_c$ between the SPT phase with the index $\pm 1 \in \mathbb{Z}_n$ and the dimer phase in the SU($n$) bilinear-biquadratic model. The broken line is a guide to the eye for the power-law decay with  $\sim n^{-5.5}$.}
\label{fig:spt_dimer}
\end{figure}

Below $\theta/\pi = -0.005$, the ground state is in the dimer phase where the entanglement spectrum changes depending on the dividing position of the system as shown in Fig.~\ref{fig:spectrum_dimer}.
The degeneracy of each level agrees with one of the dimensions of irreducible representations of
SU(4) group (shown in Appendix~\ref{appendix:dimensions}).
This type of phase transitions between an SPT phase and a trivial dimer phase was also observed in the SU(2) bilinear-biquadratic
model~\cite{su2_blbq_dimer_a,su2_blbq_dimer_b,su2_blbq_dimer_c} and the SU(3) bilinear-biquadratic
model.\cite{morimoto_ueda_momoi_furusaki} The $n$ dependence of these transition points $\theta_c$
in the SU($n$) bilinear-biquadratic models is summarized
in Fig.~\ref{fig:spt_dimer}.
They seem to show a power-law decay as a function of $n$.
We can have a naive conjecture that the transition point between the SPT and dimer phases of the SU($n$) bilinear-biquadratic model
approaches $\theta=0$ with the power law $\sim -n^{-5.5}$, which would motivate a future further study on these models.

\section{Matrix-product representation of an SU(4) VBS state with $2 \in \mathbb{Z}_4$}
\label{sec:SU4MPS}
An SU(4) symmetric VBS state with the topological index $2 \in \mathbb{Z}_4$
that is bond inversion symmetric with $\mathcal{O}^{~}_{\mathcal{I}}=0$ is most simply
constructed in the Hilbert
space of 20-dimensional single-site bases,\cite{su4_nonne_totsuka} which is naturally realized
in the two-orbital SU($n$) Hubbard model at half filling $n_g=n_e=2$.
In this section we demonstrate that the same topological class of SU(4) VBS states with
the index $2 \in \mathbb{Z}_4$ can also
appear on the 15-dimensional single-site bases at filling $n_g=3$ and $n_e=1$.
Using the MP representation, we explicitly construct the wave function and investigate
its topological properties.

\subsection{Single-site basis in the representation {\bf 15}}
To rewrite the 15-dimensional basis of three $g$-orbital fermions and one $e$-orbital fermion,
we use the fact  that the representation  \textbf{15} appears in the decomposition
of the direct product of two representations of \textbf{6},
\begin{align}
{\bf 6}\otimes\bf 6&={\bf 1}\oplus {\bf 15}\oplus {\bf 20}. \\
{\tiny \yng(1,1)}\otimes{\tiny \yng(1,1)}&={\tiny\yng(1,1,1,1)}\oplus{\tiny\yng(2,1,1)}
\oplus{\tiny\yng(2,2)} \nonumber
\end{align}
The bases of the representations \textbf{6} are made of two fermions chosen out of three $g$-orbital fermions and
one $e$-orbital fermion.
There are thus two types of six-dimensional bases:
one contains two $g$-orbital fermions,
\begin{align}
|(m,m^\prime)\rangle_{gg}=c_{g m}^\dagger c_{g m^\prime}^\dagger|0\rangle,
\end{align}
where the indexes $m,m^\prime=1,\cdots,4$ denote the nuclear spin states, and the other contains fermions with two different orbitals
($g$ and $e$),
\begin{align}
|(m,m^\prime)\rangle_{ge}={1 \over \sqrt2 }
(c_{g m}^\dagger c_{e m^\prime}^\dagger-c_{g m^\prime}^\dagger c_{e m}^\dagger)|0\rangle.
\end{align}
Hereafter we write these two sets of six-dimensional bases as
$ |{\bf 6},i\rangle_{g \alpha} $
($i=1,\cdots,6$ and $\alpha=g,e)$.

Using these two six-dimensional bases, we construct the basis in the irreducible representation
{\bf 15}
at filling $n_g=3$ and $n_e=1$ as
\begin{align}
|{\bf 15}, \sigma \rangle =
\sum_{i,i'} C^\sigma_{i,i'} |{\bf 6},i\rangle_{gg} |{\bf 6},i'\rangle_{ge}
\label{eq:15_3g1e}
\end{align}
with $\sigma=1,\dots,15$, where $C^\sigma_{i,i'}$
are the SU(4) Clebsch-Gordan
coefficients\cite{su4_clebsch_gordan} of the 15-dimensional basis in the decomposition of
$\textbf{6}\otimes \textbf{6}$.
The states (\ref{eq:15_3g1e}) are rewritten in the matrix form
\begin{align}
|{\bf 15},\sigma \rangle = {}^t {\bm u} \hat{\Gamma}^\sigma  {\bm u},
\end{align}
where
\begin{align}
{}^t {\bm u} =& (|{\bf 6},1\rangle_{gg},|{\bf 6},2\rangle_{gg},\cdots
,|{\bf 6},6\rangle_{gg},|{\bf 6},1\rangle_{ge},
\nonumber\\
&\hspace{3cm}
|{\bf 6},2\rangle_{ge},\cdots
,|{\bf 6},6\rangle_{ge}),\\
\hat{\Gamma}^\sigma =& {1 \over 2}
  \begin{bmatrix}
         \widetilde{O} & \widetilde{C}^\sigma \\
         -\widetilde{C}^\sigma & \widetilde{O} \\
  \end{bmatrix}
\end{align}
with the $6\times 6$ null matrix $\widetilde{O}$ and the antisymmetric
$6\times 6$ matrix
$\widetilde{C}^\sigma = \{ C^\sigma_{i,i'}\}$.

\subsection{Singlet bond state}
Using the two sets of six-dimensional bases in $\bm u$ as edge states, we next construct
the bond state of
two sites ($j,j+1$) where two edges are fully antisymmetrized forming four
singlet states
\begin{align}
\sum_{i,i'}C^0_{i,i'} |{\bf 6},i\rangle_{j,g\alpha} |{\bf 6},i'\rangle_{j+1,g\alpha^\prime}~,
\label{eq:1_3g1e}
\end{align}
where $C^0_{i,i'}$ are the SU(4) Clebsch-Gordan
coefficients of the singlet state in the decomposition of ${\bf 6} \otimes {\bf 6}$.
The even-parity singlet state with the bond-center inversion symmetry is
written in the matrix form
\begin{align}
{}^t {\bm u}_j \hat{\Gamma}^0_+ {\bm u}_{j+1},
\end{align}
where
\begin{align}
\hat{\Gamma}^0_+ =& {1 \over \sqrt{2+|a|^2+|b|^2}}
\begin{bmatrix}
         a \widetilde{C}^0 & \widetilde{C}^0 \\
         \widetilde{C}^0 & b \widetilde{C}^0 \\
\end{bmatrix}.
\end{align}
The ($i,i'$) element of the $6 \times 6$ matrix $\widetilde{C}^0$ is
$C^0_{i,i'}$.
The coefficients $a$ and $b$ are arbitrary, because the even-parity condition
${}^t {\bm u}_j \hat{\Gamma}^0_+ {\bm u}_{j+1} = {}^t {\bm u}_{j+1} {}^t\hat{\Gamma}^0_+ {\bm u}_{j}$ and the normalization condition $| {}^t {\bm u}_j \hat{\Gamma}^0_+ {\bm u}_{j+1} |=1$ are satisfied irrespective of $a$ and $b$.

\subsection{Matrix-product state with sixfold degenerate edge states}

Here we construct the MP state on the 15-dimensional single-site bases that have
six edge states.
To do so, we use the basis $\bm u$, which apparently has twelve edge states.

We consider the one-dimensional system with
length $L$ under the periodic boundary condition.
From the derivation of the AKLT-like ground state used in
Ref.~[\onlinecite{morimoto_ueda_momoi_furusaki}],
the MP state with bond-centered inversion symmetry is given as
\begin{align}
&{\rm Tr} \sum_{\{\sigma_j\}} (\hat{\Gamma}^{\sigma_1} \hat{\Gamma}^{0}_{+})
(\hat{\Gamma}^{\sigma_2} \hat{\Gamma}^{0}_{+}) \cdots
(\hat{\Gamma}^{\sigma_L} \hat{\Gamma}^{0}_{+})
|\sigma_1 \sigma_2 \cdots \sigma_L \rangle,\\
\label{eq:MP}
&\hat{\Gamma}^\sigma \hat{\Gamma}^0_+ ={1 \over 2\sqrt{2+|a|^2+|b|^2}}
\begin{bmatrix}
          \widetilde{C}^\sigma \widetilde{C}^0 & b \widetilde{C}^\sigma \widetilde{C}^0 \\
         - a \widetilde{C}^\sigma \widetilde{C}^0 & - \widetilde{C}^\sigma \widetilde{C}^0 \\
\end{bmatrix},
\end{align}
where $\sigma_j$ is the index of states at $j$th site ($\sigma_j=1,\cdots,15$) and
\begin{align}
|\sigma_1 \sigma_2 \cdots \sigma_L \rangle = |{\bf 15},\sigma_1\rangle_1 |{\bf 15},\sigma_2\rangle_2
\cdots |{\bf 15},\sigma_L \rangle_L .
\end{align}

We block diagonalize the $12 \times 12$ matrices $\hat{\Gamma}^\sigma \hat{\Gamma}^0_+$,
\begin{align}
U^{-1} \hat{\Gamma}^\sigma \hat{\Gamma}^0_+ U =
 \frac{\sqrt{1-a  b }}{2 \sqrt{2+|a| ^2+|b| ^2}}
\begin{bmatrix}
         \widetilde{C}^\sigma \widetilde{C}^0 & \widetilde{O} \\
         \widetilde{O} & -\widetilde{C}^\sigma \widetilde{C}^0 \\
\end{bmatrix}
\label{eq:matrix6x6_2}
\end{align}
with
\begin{align}
U=
\begin{bmatrix}
 b  \widetilde{1} & (\sqrt{1-a  b }-1)\widetilde{1} \\
 (\sqrt{1-a  b }-1)\widetilde{1} & a  \widetilde{1}
\end{bmatrix},
\end{align}
where $\widetilde{1}$ denotes the $6\times 6$ identity matrix.
Hence we can formally
rewrite the matrix product using $6 \times 6$ matrices as
\begin{align}
&{\rm Tr} (\hat{\Gamma}^{\sigma_1} \hat{\Gamma}^{0}_{+})
(\hat{\Gamma}^{\sigma_2} \hat{\Gamma}^{0}_{+}) \cdots
(\hat{\Gamma}^{\sigma_L} \hat{\Gamma}^{0}_{+})\nonumber\\
& = \! 2 \! \left( \!\! \frac{\sqrt{1-a  b }}{2 \sqrt{2+|a| ^2+|b| ^2}} \!\! \right)^{\!\!\! L}_{~}
\!\!\! {\rm Tr} (\widetilde{C}^{\sigma_1} \widetilde{C}^{0})
(\widetilde{C}^{\sigma_2} \widetilde{C}^{0}) \cdots
(\widetilde{C}^{\sigma_L} \widetilde{C}^{0}),
\end{align}
where we have assumed that $L$ is an even number.
We note that, in the product of the block diagonal matrices
Eq.~(\ref{eq:matrix6x6_2}),
two sets of six edge bases are not mixed with each other and completely decoupled.
iDMRG calculations with a restricted auxiliary space (in this case $\chi=6$ minimally) select only one of those decoupled states through the optimization process, which is the case in our numerical calculation in Fig.~\ref{fig:e_spec}.
This implies that
the (unnormalized) state is simply written as
\begin{align}
|\Psi\rangle = {\rm Tr} \sum_{\{\sigma_j\}} (\widetilde{C}^{\sigma_1} \widetilde{C}^{0})
(\widetilde{C}^{\sigma_2} \widetilde{C}^{0}) \cdots
(\widetilde{C}^{\sigma_L} \widetilde{C}^{0})
|\sigma_1 \sigma_2 \cdots \sigma_L \rangle.
\label{eq:MP2}
\end{align}
Thus only six states appear in the edge states, which are linear combinations of
$|{\bf 6},i\rangle_{gg}$ and $|{\bf 6},i\rangle_{ge}$.

We note that the matrices $\widetilde{C}^{\sigma} \widetilde{C}^{0}$ do not depend on the coefficients $a$ and $b$,
but only the edge states do.
Once we fix the coefficients $a$ and $b$ at one edge, the other edge state is also fixed,
which means the existence of a hidden long-range order.

\subsection{Symmetry operations of $Z_4 \times Z_4$ and bond-center inversion}
In the MP state (\ref{eq:MP2}),
the symmetry actions $x$ and $y$ of $Z_4 \times Z_4$ group are respectively given
by the operations of $U_x$ and $U_y$ on six-dimensional edge states in the projective representation,
\begin{align}
\sum_{\sigma'}^{} x_{\sigma,\sigma'} (\widetilde{C}^{\sigma'} \widetilde{C}^{0})
&= {U_x}^{-1} (\widetilde{C}^{\sigma} \widetilde{C}^{0}) U_x, \\
\sum_{\sigma'}^{} y_{\sigma,\sigma'} (\widetilde{C}^{\sigma'} \widetilde{C}^{0})
&= {U_y}^{-1} (\widetilde{C}^{\sigma} \widetilde{C}^{0}) U_y.
\end{align}
From straightforward calculations, we find that
these two satisfy $U_x^4=U_y^4=\tilde{1}$ and
\begin{align}
U_x U_y = -U_y U_x.
\label{eq:comm}
\end{align}
This shows that the state (\ref{eq:MP2})
belongs to the $\mathbb{Z}_4$ SPT phase with the topological index $2\in \mathbb{Z}_4$.

This state has the bond-center inversion symmetry.
As for the relation of Eq.~(\ref{eq:inversion}) in Appendix~\ref{app:inversion} for this state, we obtain
\begin{equation}
{}^t(\widetilde{C}^{\sigma} \widetilde{C}^{0}) = - (\sqrt{6}\widetilde{C}^0_{~})^{-1}_{~}
 (\widetilde{C}^{\sigma} \widetilde{C}^{0}) (\sqrt{6}\widetilde{C}^0_{~}),
\end{equation}
where we have used the relations ${}^t\widetilde{C}^0=\widetilde{C}^0_{~}$, ${}^t\widetilde{C}^\sigma=-\widetilde{C}^\sigma_{~}$, and $(\sqrt{6}\widetilde{C}^0_{~})^{2}_{~}=\widetilde{1}$. This shows that the state has the index $\mathcal{O}^{~}_{\mathcal{I}}=0$,
where $\theta^{~}_{\mathcal{I}} = \pi$ and $U^{~}_{\mathcal{I}}=\sqrt{6}\widetilde{C}^0_{~}$.
This state also obviously has the translational symmetry.

We thus find that the MP state (\ref{eq:MP2}) belongs to the same topological class as the ground state of the SU(4) $e$-$g$ spin model shown in
Sec.~\ref{sec:model} and also the VBS state\cite{su4_nonne_totsuka} studied at half filling $n_g=n_e=2$.

\subsection{Correlations}\label{sec:SU4MPS_corr}
The bond states of two sites are projected to the irreducible representations
 ${\bf 20}$, ${\bf 15}$, and ${\bf 1}$.
The quadratic Casimir operator on the bond takes the value
\begin{eqnarray}
\left \langle \big( {\bm T}_j + {\bm T}_{j+1} \big)^2 \right\rangle_{2 \in \mathbb{Z}_4} & = & 24/5,
\end{eqnarray}
whereas, in the case of the $\mathbb{Z}_4$ SPT phase with the index $\pm 1 \in \mathbb{Z}_4$,\cite{morimoto_ueda_momoi_furusaki}
it takes
\begin{eqnarray}
\left \langle \big( {\bm T}_j + {\bm T}_{j+1} \big)^2 \right\rangle_{\pm1 \in \mathbb{Z}_4} & = & 56/15,
\end{eqnarray}
where the bond states are projected only to ${\bf 15}$ and ${\bf 1}$.
We can understand the difference of the values of the Casimir operator from the fact
that a representation ${\bf p}$
with a larger dimension takes a larger eigenvalue $C({\bf p})$ of the Casimir operator, e.g.,
 $C({\bf 1}) = 0$, $C({\bf 15}) = 4$, and $C({\bf 20}) = 6$.

The correlation length of the MP states can be obtained from the largest and second largest eigenvalues $\epsilon_1$ and $\epsilon_2$
(in terms of absolute values) of the transfer matrix $\sum_{\sigma} (\widetilde{C}^{\sigma} \widetilde{C}^{0})^* \otimes (\widetilde{C}^{\sigma} \widetilde{C}^{0})$ as
\begin{equation}
\xi = - 1 / \ln|\epsilon_2/\epsilon_1|.
\end{equation}
For the MP state Eq. (\ref{eq:MP2}) with $2 \in \mathbb{Z}_4$,
this procedure gives
the correlation length $\xi = 1/\ln 5$.
We note that the correlation length of the MP state with
$\pm 1 \in \mathbb{Z}_4$ is reported as $\xi =1/\ln 15$.~\cite{Rachel2010} 

\section{Summary and discussions}
\label{sec:summary}
\begin{figure*}
\includegraphics[width=13cm]{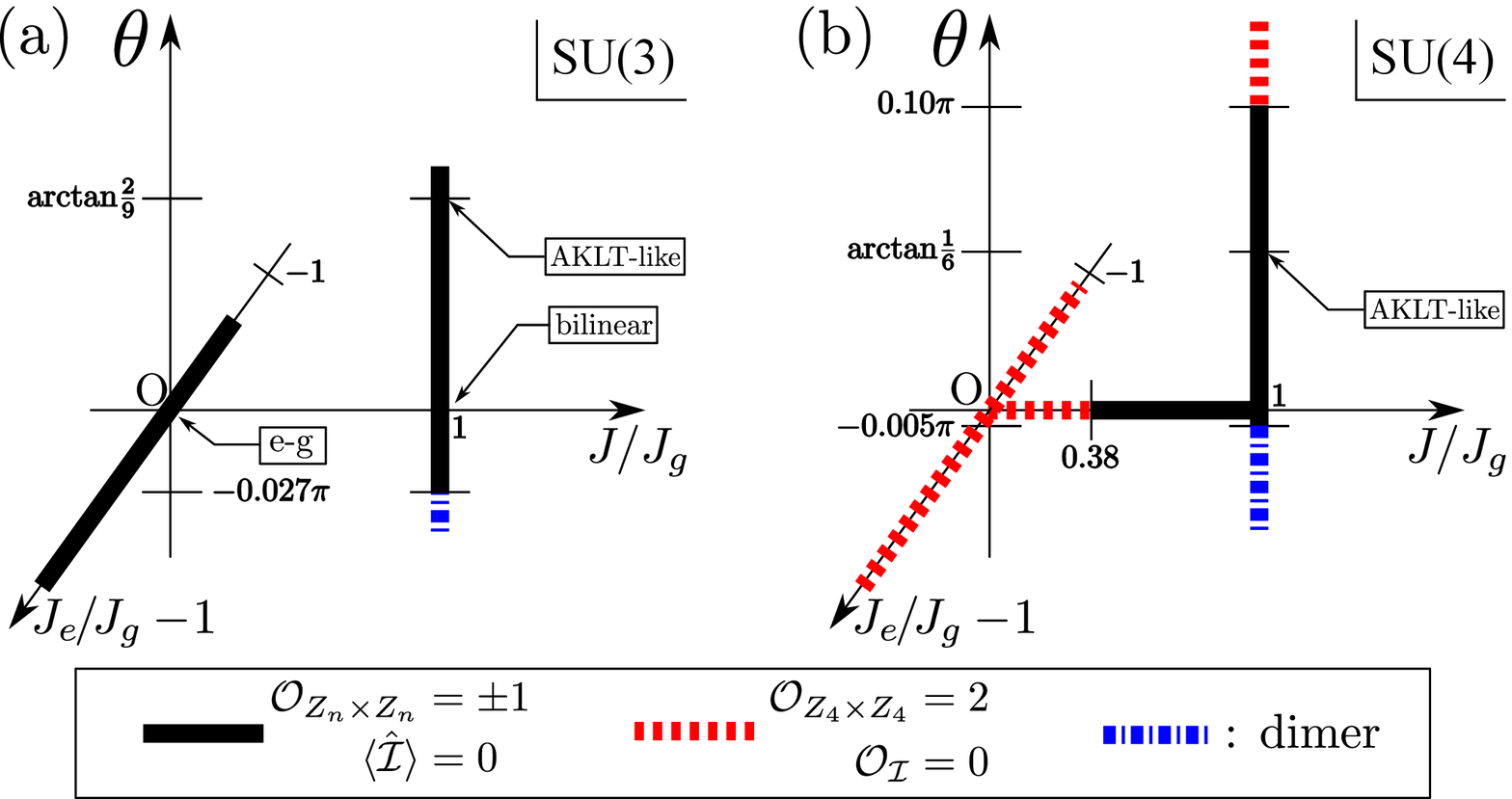}
\caption{(Color online) Phase diagrams for SU($n$) spin models of (a) $n=3$ and (b) $n=4$.
The thick  lines denote the
VBS phases with the topological index $\pm 1 \in \mathbb{Z}_n$ and the dotted (red) lines the VBS phases
with
$2 \in \mathbb{Z}_4$. The dotted dash (blue) lines denote dimer phases.}
\label{fig:phase}
\end{figure*}

We have studied topological properties of the ground state of the one-dimensional SU($n$)
($n=3$ and 4) spin model called the SU($n$) $e$-$g$ spin model, which captures low-energy physics of two-orbital SU($n$)
fermionic-atom systems with the filling of $n_g=n-1$ ground-state atoms and $n_e=1$ excited-state atoms.
We have also investigated the phase diagrams in the parameter space that includes the SU($n$) $e$-$g$ spin model, the SU($n$) bilinear model, and also the SU($n$) bilinear-biquadratic model on the local bases
of the ($n^2-1$)-dimensional adjoint representation of SU($n$) group.
The obtained phase diagrams are summarized in Fig.~\ref{fig:phase}.

Our iDMRG results show that the ground states of the SU($n$) $e$-$g$ spin models are in the nontrivial $\mathbb{Z}_n$ SPT phases with topological indices $\pm1 \in \mathbb{Z}_3$ for $n=3$ and $2 \in \mathbb{Z}_4$ for $n=4$, which are protected by $Z_n \times Z_n$ symmetry. In the case of $n=4$, the ground state is topologically distinct from the ground state of the SU(4) bilinear model in the 15-dimensional
representation, whereas, in the SU(3) symmetric case,
the ground state is topologically identical with the ground state of the SU(3) bilinear model
in the eight-dimensional representation.
Thus the SU(4) spin system in the adjoint representation {\bf 15} contains three distinct  nontrivial $\mathbb{Z}_4$ SPT phases  in the phase diagram.
A phase transition occurs between the $\mathbb{Z}_4$ SPT phase with $2 \in \mathbb{Z}_4$ and that with $\pm1 \in \mathbb{Z}_4$, where the bond-inversion symmetry breaking occurs simultaneously.
An SPT phase transition of the same type appears in the SU$(4)$ bilinear-biquadratic model (\ref{eq:H_BBQ}) when the  magnitude of the biquadratic interaction
is given by $\theta \sim 0.10\pi$. This variation of interactions controls the antisymmetrization effect on bond degrees of freedom,
thereby changing the edge degrees of freedom. We capture this behavior in the increase of the expectation value of the quadratic
bond-Casimir operator at the $\mathbb{Z}_4$ SPT phase transition.

We have demonstrated how to construct a valence bond solid state with the topological index
$2 \in \mathbb{Z}_4$ in the 15-dimensional representation (equivalent to the case $n_g=3$, $n_e=1$) and shown that this state is
topologically identical with the SU(4) AKLT state\cite{su4_nonne_totsuka} at half filling $n_g=n_e=2$.
In the case of SU(4) symmetric one-dimensional chains, there are three nontrivial SPT phases.
On the 15-dimensional representation bases,
we can write down all of three different SU(4) VBS states in the MP representation which
belong to distinct $\mathbb{Z}_4$ SPT phases;
the MP states with the index
$\pm1 \in \mathbb{Z}_4$ are studied in Ref.~[\onlinecite{morimoto_ueda_momoi_furusaki}]
and that with $2 \in \mathbb{Z}_4$ is shown in Sec.~\ref{sec:SU4MPS}.

We note that a similar analysis was performed in Ref.~[\onlinecite{roy2018}] by using a
diagrammatic method. They obtained  the SU($n$) AKLT states with the indexes
$\pm 1 \in \mathbb{Z}_n$ and $\pm 2 \in \mathbb{Z}_n$ on the
($n^2-1$)-dimensional adjoint representation and presented the explicit forms of the Hamiltonians. For SU(4), these two classes correspond
to the classes $\pm 1 \in \mathbb{Z}_4$ and $2 \in \mathbb{Z}_4$.  Their AKLT state in $2 \in \mathbb{Z}_4$ is essentially the same one as ours.

An interesting future issue would be proposals for realizations of the cross coupling $J$ and the biquadratic interactions in the cold atom systems that are necessary to observe the SPT phase transitions experimentally. Another future issue is to understand the properties of the SU($n$) spin models in the large-$n$ limit and understand the behavior of the SU($n$) models with higher $n$ systematically, which is hard to treat by non-perturbative approaches.
As discussed in Sec. \ref{sec:su_n_blbq}, we expect the existence of the phase transition between the SPT phase with $\pm1 \in Z_n$ and the dimer phase in the SU($n$) bilinear-biquadratic model for general $n$, where we conjecture that the transition point $\theta_c$ approaches $\theta=0$ with the power law of $n$.
Moreover, there are more choices of filling pattern in the two-orbital SU($n$) model for larger $n$. Thus we expect that the ground-state phase diagrams show more structures depending on the choice of the filling.
A promising numerical approach for the investigation of this problem is an extended
non-Abelian DMRG adapted for the SU($n$) symmetry.\cite{nataf_mila_PRB2018, Weichselbaum2018}

\section*{Acknowledgments}
The authors thank Akira Furusaki and Keisuke Totsuka for fruitful discussions.
The work was partially supported by Grants-in-Aid for Scientific Research (KAKENHI) under Grants No.~JP25800221, No.~JP17K14359 (H.U.),
and No.~JP16K05425 (T.M.) from Japan Society for the Promotion of Science (JSPS) and
the Gordon and Betty Moore Foundation's EPiQS Initiative Theory Center Grant (T.M.).

\appendix
\section{DIMENSIONS OF THE IRREDUCIBLE REPRESENTATIONS}
\label{appendix:dimensions}

In this appendix, we list the dimensions of the irreducible representations of both SU(3)
and SU(4) symmetry groups.
For the SU(3) group,
the irreducible representations are described with the Young tableau as
\begin{eqnarray}
\overbrace{\yng(3,3)}^{m} \hspace{-1.9mm} \raisebox{5.6pt}{ $\overbrace{\yng(2)}^{n}$ }, \nonumber
\end{eqnarray}
where $n,m=0,1,2,\cdots$. The dimensions of these representations are given by~\cite{table_su3}
\begin{eqnarray}
D^{nm}
& = & \frac{1}{2}(n+1)(m+1)(n+m+2).
\label{eq:dim_irrep_su3}
\end{eqnarray}
For the SU(4) group, the irreducible representations are described as
\begin{eqnarray}
\overbrace{\yng(4,4,4)}^{q} \hspace{-1.8mm} \raisebox{5.6pt}{ $\overbrace{\yng(3,3)}^{r}$ } \hspace{-2.4mm} \raisebox{11.2pt}{ $\overbrace{\yng(2)}^{p}$ } ,\nonumber
\end{eqnarray}
where $q,r,p=0,1,2,\cdots$.
The dimensions of them are given by
\begin{eqnarray}
D^{prq}
& = & \frac{1}{12}(p+1)(q+1)(r+1)(p+r+2) \nonumber \\
&  & \times (q+r+2)(p+q+r+3) .
\label{eq:dim_irrep_su4}
\end{eqnarray}
We show the dimensions of the first several irreducible representations $D^{mn}$ and $D^{prq}$
in Table.~\ref{table:dim_irrep}.

\begin{table}[tb]
\begin{ruledtabular}
  \caption{Dimensions of first several irreducible representations $D^{mn}$ for the SU(3) group
  given by Eq.~(\ref{eq:dim_irrep_su3}) and $D^{prq}$ for the SU(4) given by Eq.~(\ref{eq:dim_irrep_su4}).}
  \label{table:dim_irrep}
  \begin{tabular}{cc|cc}
    set of $D^{nm}$ & dimension & set of $D^{prq}$ & dimension  \\
\hline
$D^{00}$ & 1 & $D^{000}$ & 1 \\
$D^{10}$ & 3 & $D^{100}$ & 4  \\
$D^{20}$ & 6 & $D^{010}$ & 6 \\
$D^{11}$ & 8 & $D^{101}$ & 15 \\
$D^{30}$ & 10 & $D^{020}$, $D^{110}$ & 20 \\
$D^{21}$ & 15 & $D^{201}$ & 36 \\
$D^{31}$ & 24 & $D^{111}$ & 64 \\
  \end{tabular}
\end{ruledtabular}
\end{table}

\section{TOPOLOGICAL INDEX ASSOCIATED WITH INVERSION SYMMETRY \label{app:inversion}}
In this section we briefly describe how to obtain $\mathcal{O}_{\mathcal{I}}$ numerically.
In the MP representation,
the bond-center inversion $\hat{\mathcal{I}}$ transforms as
\begin{equation}
\hat{\mathcal{I}} \ket{\Psi} = \sum_{\{ \sigma_j \}} \Tr \prod_j {}^t\! A^{\sigma_j} \ket{\sigma_1 \sigma_2 \cdots \sigma_L}~.
\end{equation}
When $\ket{\Psi}$ has the inversion symmetry, the transformation law in the projective
representation is given by
\begin{equation}
{}^t\!A^{\sigma_j} = e^{i \theta_{\mathcal{I}}} U^{-1}_\mathcal{I} A^{\sigma_j} U^{~}_\mathcal{I}
\label{eq:inversion}
\end{equation}
with a phase $\theta_{\mathcal{I}} \in \{0,\pi \}$
and a $\chi \times \chi$ matrix $U_\mathcal{I}$.
Applying the inversion operation  again, we obtain
\begin{equation}
A^{\sigma_j} = e^{i2\theta_{\mathcal{I}}} {}^tU^{~}_{\mathcal{I}} U^{-1}_{\mathcal{I}} A^{\sigma_j} U_{\mathcal{I}} {}^tU^{-1}_{\mathcal{I}},
\end{equation}
which is reduced to the relation~\cite{spt_pollmann_oshikawa}
\begin{equation}
{}^tU^{~}_{\mathcal{I}} = e^{i \pi \mathcal{O}_{\mathcal{I}}} U_{\mathcal{I}}
\end{equation}
with the index $\mathcal{O}_{\mathcal{I}} \in \{0,1\}$.
We can thus classify each inversion symmetric phase with the discrete index $\mathcal{O}_{\mathcal{I}}$.
To extract $\mathcal{O}_{\mathcal{I}}$, we calculate
\begin{equation}
\mathcal{O}_{\mathcal{I}} = \frac{1}{\pi i } \log \left[ \frac{1}{\chi} \Tr \big\{ {}^tU_{\mathcal{I}} U^{-1}_{\mathcal{I}} \big\} \right]~.
\end{equation}

\bibliographystyle{apsrev4-1}
%

\end{document}